\documentclass[aps,prx,twocolumn,superscriptaddress,longbibliography,floatfix,showpacs]{revtex4-1}
\usepackage{amsmath}
\usepackage{amsfonts}
\usepackage{amssymb}
\usepackage{graphicx}
\usepackage{bm}
\usepackage{float}
\usepackage{comment}

\begin{document}

\title{Indirect Spin Exchange Interaction in Substituted Copper Phthalocyanine Crystalline Thin Films}
\author{Naveen Rawat}
\email{Naveen.Rawat@uvm.edu}
\affiliation{Department of Physics and the Materials Science Program, University of Vermont, Burlington, Vermont 05405, USA}
\author{Zhenwen Pan} 
\affiliation{Department of Physics and the Materials Science Program, University of Vermont, Burlington, Vermont 05405, USA}
\author{Cody J. Lamarche}
\affiliation{Department of Physics and the Materials Science Program, University of Vermont, Burlington, Vermont 05405, USA}
\author{Takahisa Tokumoto}
\affiliation{National High Magnetic Field Laboratory, Tallahassee, FL 32310, USA}
\author{Judy G. Cherian}
\affiliation{National High Magnetic Field Laboratory, Tallahassee, FL 32310, USA}
\author{Anthony Wetherby}
\affiliation{Department of Chemistry, University of Vermont, Burlington, Vermont 05405, USA}
\author{Rory Waterman}
\affiliation{Department of Chemistry, University of Vermont, Burlington, Vermont 05405, USA}
\author{Randall L. Headrick}
\affiliation{Department of Physics and the Materials Science Program, University of Vermont, Burlington, Vermont 05405, USA}
\author{Stephen A. McGill}
\affiliation{National High Magnetic Field Laboratory, Tallahassee, FL 32310, USA}
\author{Madalina I. Furis}

\email{Corresponding author : Madalina.Furis@uvm.edu}
\affiliation{Department of Physics and the Materials Science Program, University of Vermont, Burlington, Vermont 05405, USA}
\begin{abstract}
The origins of indirect spin exchange in crystalline thin films of Copper Octabutoxy Phthalocyanine (Cu-OBPc) are investigated using Magnetic Circular Dichroism (MCD) spectroscopy. These studies are made possible by a solution deposition technique which produces highly ordered films with macroscopic grain sizes suitable for optical studies. For temperatures lower than 2 K, the contribution of a specific state in the valence band manifold originating from the hybridized lone pair in nitrogen orbitals of the Phthalocyanine ring, bears the Brillouin-like signature of an exchange interaction with the localized \textit{d}-shell Cu spins. A comprehensive MCD spectral analysis coupled with a molecular field model of a $\sigma\pi-d$ exchange analogous to \textit{sp-d} interactions in Diluted Magnetic Semiconductors (DMS) renders an enhanced Zeeman splitting and a modified \textit{g}-factor of -4 for the electrons that mediate the interaction. These studies define an experimental tool for identifying electronic states involved in spin-dependent exchange interactions in organic materials.
\end{abstract}
\date{\today}
\maketitle

\section{\label{sec:int} Introduction}
\indent Metal phthalocyanines (MPcs) are often presented as the archetype of small molecule organic semiconductors because they are employed in the vast majority of organic optoelectronics applications \cite{Walter2010,Sergeyev2007,Siebbeles2009,Najafov2010,Hains2010}. These semiconductors form an interesting class of materials whose electronic and optical properties are the result of an interplay between localization and delocalization of spin and charge carriers. 3-\textit{d} transition MPcs are fully compatible with all organic devices processing techniques and, most importantly, the presence of the metal ion results in a potential for customization of electronic properties \cite{Deboer2005,Zeis2005,Tang2006}. Despite the significant body of research work that explores the electronic properties of these molecules, magnetism studies of MPcs in the crystalline phase are especially rare and raise numerous challenges with regards to unambiguously identifying the exchange mechanisms and electronic states involved in the observed collective magnetic state \cite{Ishikawa2010,Bartolome2014}. Theoretical models predict that in crystalline 3\textit{d}-transition MPc thin films, a spin-dependent exchange interaction exists between the delocalized $\pi$-electrons of the phthalocyanine C-N ring and the localized unpaired \textit{d}-shell electron spin of the central metal ion \cite{Bartolome2014,Grobosch2010,Wu2008,Marom2008,Kroll2012}. The exchange mechanisms are predicted to be very different depending on the molecular ordering in the crystalline phase, the number of unpaired spins on the \textit{d}-orbitals of the central ion, the energy and symmetry of these orbitals relative to those of the ligand $\pi$-electrons. Elucidating these exchange mechanisms is very important for magneto-optics and spintronics applications \cite{Wolf2001,Zutic2004}, especially in the context of the great progress made in recent years on developing crystalline small molecule thin films with very large electron mobilities \cite{Diao2013,Takeya2007,Wang2012}. One can therefore envision the MPcs as organic analogues for the Diluted Magnetic Semiconductors (DMS), a system where conduction/delocalized electrons are spin polarized through an exchange interaction with unpaired, localized \textit{d}-shell electrons of a magnetic impurity \cite{Dietl2014,Dietl2008,Ohno2013,Ohno2010,Furdyna1988}.

A comprehensive study of exchange interactions in these materials must necessarily start with copper phthalocyanine (CuPc), the most studied and best understood member of this family in terms of its electronic and magnetic properties \cite{Lee1987,Heutz2007,Lozzi2004,Oteyza2010,Willey2013}. The $d^{\textit{9}}$ configuration of the Cu$^{2+}$ ion results in a single unpaired electron spin located in a 3$d_{x^{2}-y^{2}}$ or $b_{1g}$ orbital. Theoretical studies predict that this spin engages in a Coulomb-like indirect exchange interaction with the delocalized $\pi$-electrons of the ligand, resulting in a collective anti-ferromagnetic ($\alpha$-phase) or weakly ferromagnetic ($\beta$-phase) ordering at low temperatures in crystalline thin films \cite{Heutz2007,Kroll2012,Wu2011}. Magnetization measurements do confirm the presence of long range magnetic ordering and finite ordering temperatures, but they cannot directly identify the electronic state(s) mediating this indirect exchange, or measure the strength of this exchange mechanism between the \textit{d}-shell unpaired spins and the delocalized ligand electrons. This is a consequence of the thermodynamic nature of SQUID magnetometry that only provides information on the overall interaction between unpaired spins, but it is unable to resolve the contributions of conduction/delocalized electrons to this mechanism. In contrast, polarization -resolved magneto-spectroscopy techniques, such as Magnetic Circular Dichroism (MCD) can be used to investigate the nature of the exchange mechanisms because they resolve the interaction between \textit{d}-shell ions and each electronic state in the valence band manifold, an interaction which results in spin polarization of delocalized electrons while facilitating the \textit{d}-shell ion spin ordering \cite{Tiedemann2011,Ando2006,Ostrovsky2011,Xavier2009}.

In this research article, we report the results of MCD spectroscopy experiments performed on crystalline thin films of copper(II) 1,4,8,11,15,18,22,25-octabutoxy-29H,31H-phthalocyanine (Cu-OBPc) with macroscopic grain sizes. Details about the samples and the principles of MCD measurements are presented in Sec.\ref{sec:mcd}. The magnetic field and temperature dependent MCD presented in Sec.\ref{sec:three} unambiguously identifies the electronic state within the HOMO-LUMO bandgap manifold mediating the indirect exchange between the Cu ions. A quantitative estimation for the strength of Coulomb exchange interactions between the \textit{d}-shell spins and the ligand electrons that occupy this state is also obtained in the form of an enhanced effective \textit{g}-factor for the ligand electrons, as discussed in Sec.\ref{sec:aed}. Our understanding of the indirect exchange interaction that leads to this \textit{g}-factor enhancement is also presented in Sec.\ref{sec:aed}. The solution processing method employed to fabricate the films is rectangular capillary hollow pen writing technique \cite{Headrick2008}, that routinely produces high mobility small molecule thin films with mm-sized grains, on transparent substrates, as described in Sec.\ref{sec:exp}. 

\section{\label{sec:mcd} MCD of Solution-Processed Crystalline Thin Films}
\indent Since the first optical rotation measurements performed by Michael Faraday \cite{Faraday1846}, spectroscopy experiments continuously evolved to explore different aspects of magnetic phenomena in a broad array of materials ranging from heme-proteins to DMSs \cite{Ranjbar2009,Bussian2009,Ando2006}. One of these experiments is Magnetic Circular Dichroism (MCD), that measures the differences in absorbance between right ($\sigma^{+}$) and left ($\sigma^{-}$) circularly polarized light in a sample ($ MCD \sim (A_{RCP}-A_{LCP}$) in the presence of a DC magnetic field parallel to the \textbf{k}-vector of incident light(Faraday geometry). This difference is a measure of the change in angular momentum projection on the applied magnetic field direction for electrons that absorb the circularly polarized photons in accordance with conservation of angular momentum. Unlike calorimetry or SQUID-based magnetization that probe the statistical ensemble average of all contributions to the net magnetization, MCD directly probes the exchange mechanism responsible for the macroscopic magnetization by spectrally resolving the net total angular momentum of individual excited states. In the presence of spin-orbit coupling, this is a measure of the electron spin polarization.  Furthermore, constant advancements in high magnetic fields/low temperature technologies and state-of the art polarization optics manufacturing techniques enable MCD experiments in small molecule organic semiconductors where spin-exchange and spin-orbit coupling are orders of magnitude weaker than typical magnetic thin films \cite{Gredig2012,Pan2012,Pan2011}.

\begin{figure}[h!]
	\centering
	\includegraphics[width=\columnwidth]{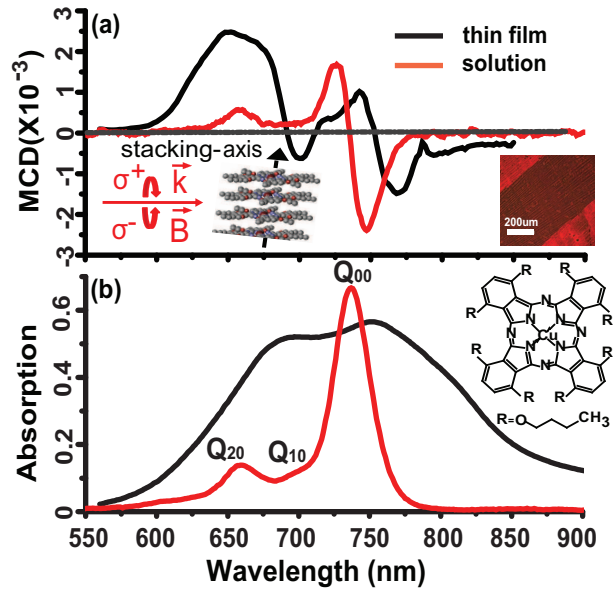}
	\caption{\label{fig:1} (Color online) (a) MCD spectra of Cu-OBPc molecules in toluene solution measured at B = 1 T and T = 300 K (red curve) and a Cu-OBPc polycrystalline thin film recorded at B = 10 T and T = 0.4 K (black curve)Inset (right): Polarized microscope image of the Cu-OBPc thin film; scale bar = 200 $\mu$m Inset (left): experimental geometry showing the stacking axis is perpendicular to the $\vec{k}$-vector of incident, right ($\sigma^{+}$) or left ($\sigma^{-}$) circularly polarized light in the Faraday geometry ($\vec{k}\parallel\vec{B}$)  (b) Absorption spectra of Cu-OBPc molecules in toluene solution (red curve) and a Cu-OBPc polycrystalline thin film (black curve) recorded in the Q-band spectral region. Inset: Cu-OBPc molecule. Experimental details stated in Sec.\ref{sec:exp} }
\end{figure}

The MCD study reported here employs an organo-soluble copper phthalocyanine (Cu-OBPc), where the octabutoxy groups substituted in the non-peripheral positions of the phthalocyanine ring improve the solubility of CuPc in common organic solvents, enabling solution processing techniques for thin film fabrication \cite{Cook1987,Kobayashi2003}. Solution processing techniques are quite successful in producing long range ordered thin films with a significantly reduced number of grain boundaries and minimal defects and disorder, a crucial factor in improving electronic properties and device performance \cite{Rivnay2009,Abe2014,Wo2012}. The molecule of choice, Cu-OBPc, was purified by recrystallization and crystalline thin films were fabricated from a toluene solution (see Sec.\ref{sec:exp} for details) on \textbf{c}-plane (MCD - inactive) crystalline sapphire substrates using the rectangular capillary hollow pen writing technique \cite{Rawat2015,Wo2012,Cour2013,Headrick2008}. A typical polarized microscope image of the resulting Cu-OBPc film is shown in  Fig. 1(a) inset where the scale bar is 200 $\mu$m. The image contrast originates in the different orientation of the crystalline axes in neighboring grains. In crystalline Cu-OBPc, molecules pack tightly along a particular direction further referred to as the stacking axis \cite{Zhang2009}. Selection rules dictate that light polarized along this axis is not absorbed \cite{Davydov1971}, giving rise to the dark central region of the inset image, where the stacking axis happens to be oriented parallel to the microscope polarizer axis. At this point, it is important to note that the high mobility axis also coincides with the stacking axis and all significant orbital overlap and long range interactions between neighboring molecules happen along this same axis. For this reason phthalocyanines are often referred to as quasi-1D systems.

\begin{figure}
\centering	
\includegraphics [width = \columnwidth]{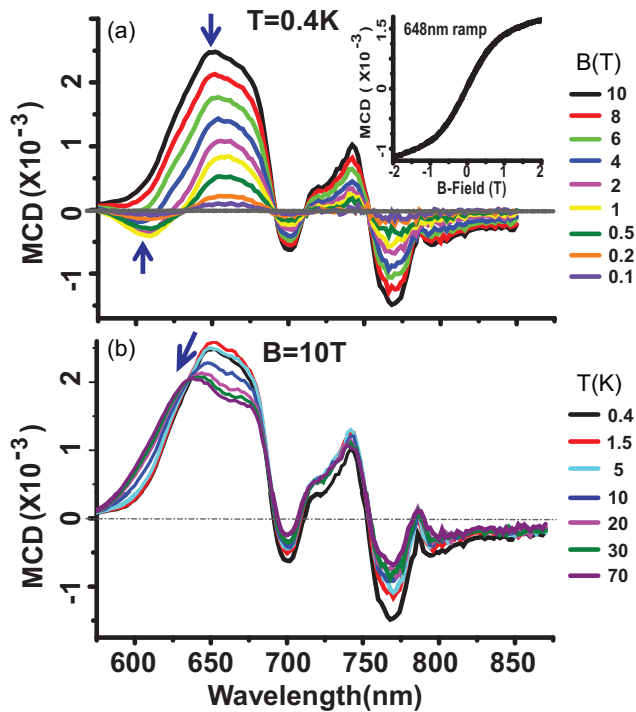}
\caption{\label{fig:2} (Color online) (a) MCD spectra of Cu-OBPc polycrystalline thin film recorded at T = 0.4 K and varying magnetic fields up to 10 T. Inset: MCD vs. B recorded at $ \lambda $ = 648 nm during a magnetic field ramp (b) MCD spectra from the same film recorded at B = 10 T and varying temperatures up to 70 K.}
\end{figure}

\indent MCD and absorption measurements were carried out with the sample mounted in a 10 T Oxford superconducting  magnet (Spectromag) equipped with a He3 insert for reaching sub-kelvin sample temperatures. In these crystalline thin films, the molecules stack edge-on with their stacking axis parallel to the substrate. The inset in Fig. 1(a) also illustrates the orientation of the magnetic field with respect to the stacking axis and the incident light. More details about the experimental setup can be found in Sec.\ref{sec:exp}. Figure 1(a) shows a typical MCD spectrum of the Cu-OBPc Q-band region in solution (red curve) and crystalline thin film (black curve) while Fig. 1(b) displays the respective absorption spectra. For the thin film, the measurement was carried out at 10 Tesla (T) and 0.4 Kelvin (K), whereas the solution MCD was measured at 1 T and room temperature using an electromagnet and scaled in intensity for a direct comparison. In the solution spectrum, we can easily identify and assign three MCD-active Q band transitions based on well-known results available in literature \cite{Huang1981,Keizer2003,Nyokong1987}. Q$_{00}$ at 736 nm corresponds to $\pi$-$\pi^{*}$ ($a_{1u}$ to $e_{g}$) transition at the HOMO-LUMO gap and is polarized in the plane of the molecule. Q$_{10}$ at 696 nm is the contribution of vibrational overtones and Q$_{20}$ located at 658 nm can be assigned as a transition from lower lying $\sigma$ states in the valence band manifold ($e_{u}(\sigma) \rightarrow e_{g}(\pi^{*}$)), polarized perpendicular to the molecular plane and vibronically coupled to Q$_{00}$ \cite{Dunford1997}. Absorption spectrum of the Cu-OBPc thin films is significantly broadened compared to that of the solution due to the Davydov splitting and excitonic coupling in the crystalline phase that leads to more complex band structure \cite{Davydov1971,Hollebone1978}. The most striking feature of the thin film MCD spectrum is the significant increase in the MCD magnitude in the Q$_{20}$ spectral region. The $\pi$ orbital overlap and the phonon coupling leads to a degeneracy lifting of the electronic states, resulting in a manifold of MCD-active states in that spectral region. An extensive discussion of MCD spectral fitting and the assignment of these states is included in Sec.\ref{sec:aed}

\section{\label{sec:three} MCD Signatures of Low Temperature Magnetic Ordering in Cu-OBPc}
\indent  Figure 2(a) displays a series of MCD spectra recorded in a Cu-OBPc thin film at 0.4 K for different magnetic fields ranging from 0.1 T to 10 T. While the spectra contains contributions from all the states in the bandgap manifold, only the MCD associated with the $e_{u}$($\sigma$) to $e_{g}$($\pi^{*}$) transitions (marked with blue arrows in the figure) evolves with magnetic field and temperature in a non-trivial way. The MCD intensity recorded during a magnetic field sweep at 648 nm (the wavelength corresponding to the largest MCD signal in the aforementioned spectral region) displays a marked non-linear evolution with magnetic field (see Fig. 2(a) inset) indicative of a net spin polarization. The temperature evolution of the B = 10 T MCD spectrum presented in Fig. 2(b) confirms the distinctly singular evolution of the 648 nm feature. The contribution to the MCD spectrum from this state blueshifts and decreases in intensity with increasing temperature, in contrast with the rest of the spectrum which is weakly temperature dependent, with only a small decrease between 0.4 K and 5 K and is not spectrally shifted. 
 
In order to confirm the distinct evolution of the 648 nm contribution with magnetic field, the MCD values extracted from the spectra in Fig. 2 were plotted as a function of magnetic field at different temperatures in Fig. 3(a). MCD values from the 1.5 K spectrum of a non-magnetic Zn-OBPc thin film were also included for comparison. The complete MCD spectrum for Zn-OBPc presented in Figure S1 of the Supporting Information section of this paper \cite{SuppInfo}. The 100 K and 300 K MCD vs. B field sweeps were recorded using an exact replica of the MCD setup that was re-built in cell 5 of the National High Magnetic Field Laboratory (NHMFL) for the 25 T Split-Florida Helix magnet \cite{Pan2011}. At high temperatures, the MCD vs. B curve is strictly linear, a signature of the expected, large diamagnetic, contribution to the MCD spectra, universally present in organic systems where orbital momentum quenching is less effective \cite{Tiedemann2011,Kobayashi2007,Mack2007}. At lower temperatures (see Fig. 3(a)), the 648 nm MCD in Cu-OBPc increases at a much steeper slope, an indication of the onset of exchange interaction between the \textit{d}-shell unpaired spin of Cu$^{2+}$ and the delocalized bandgap electrons spins. This type of behavior has been previously observed for electrons in DMSs (\textit{e. g}. CdSe and ZnSe) doped with magnetic \textit{d}-shell ions, in particular Mn, Fe or Cr \cite{Hwang2013,Dagnelund2012,Bussian2009,Bacher2001,Titova2002}. In that case the MCD vs. B curve followed a Brillouin-like function, saturating at high fields where all \textit{d}-shell spins are aligned with the applied magnetic field. In the case of Cu-OBPc the MCD never saturates but rather continues to increase linearly with magnetic field at a slope identical to the diamagnetic contribution. In contrast, the MCD of Zn-OBPc at 1.5 K is almost identical to that of Cu-OBPc at 70 K. Since the two molecules have identical ligands and their valence band density of states is virtually identical, as evidenced by electron energy-loss spectroscopy \cite{Grobosch2010}, it is not surprising that the diamagnetic behavior is identical for the two films. It is also very strong evidence that in the case of Cu-OBPc, the Brillouin -like behavior superimposed on the diamagnetism is not originating from the ligand states alone, but rather from the exchange between the \textit{d}-shell localized states and the ligand, delocalized $\pi$ - electrons. In fact, low temperature MCD studies on isolated CuPc molecules in an argon matrix \cite{Dunford1997} show a temperature independent, entirely different behavior for all the three transitions identified in the single molecule spectrum in Fig. 1(a). The temperature dependent MCD evolution implies a magnetic phase transition is happening in Cu-OBPc from diamagnetism at high temperatures to an ordered state at low temperatures, where the exchange mechanism between \textit{d}-shell and $\pi$ electrons is also present.

The temperature dependence in Fig. 2(b) is consistent with previous SQUID magnetization studies in the unsubstituted CuPc molecule \cite{Heutz2007}. Figure 3(b) summarizes the evolution of the maximum MCD associated with this $e_{u}(\sigma) \rightarrow e_{g}(\pi^{*}$) transition and the corresponding peak energy as a function of temperature at B = 10 T. Since the MCD vs. T curve in our experiment  can be interpreted similarly to the temperature dependence of a magnetic susceptibility, the initial increase in MCD from 0.4 K to 1.5 K, followed by a sharp decay is attributed to the antiferromagnetic ordering below 2 K. The MCD peak energy (marked with an arrow in Fig. 2(b)) blueshifts with temperature, remaining constant for T $ > $ 50 K. This is the very same temperature where the MCD spectra become temperature-independent similar to those of a diamagnetic system (100K and 300K MCD from this sample can be found in \cite{Pan2011}). It is also important to note that the high energy side of the low magnetic field MCD spectrum switches from a derivative-like shape characteristic for degenerate ground states to a gaussian-like shape associated with a non-degenerate state at the same temperature. We can therefore conclude that the spin-orbit coupling turns on at 50 K for this particular state in the valence band manifold, enabling the onset of ordering of Cu spins through an indirect exchange mechanism around the same temperature. 
\begin{figure}[h!]
	\centering
	\includegraphics[width=8.6cm]{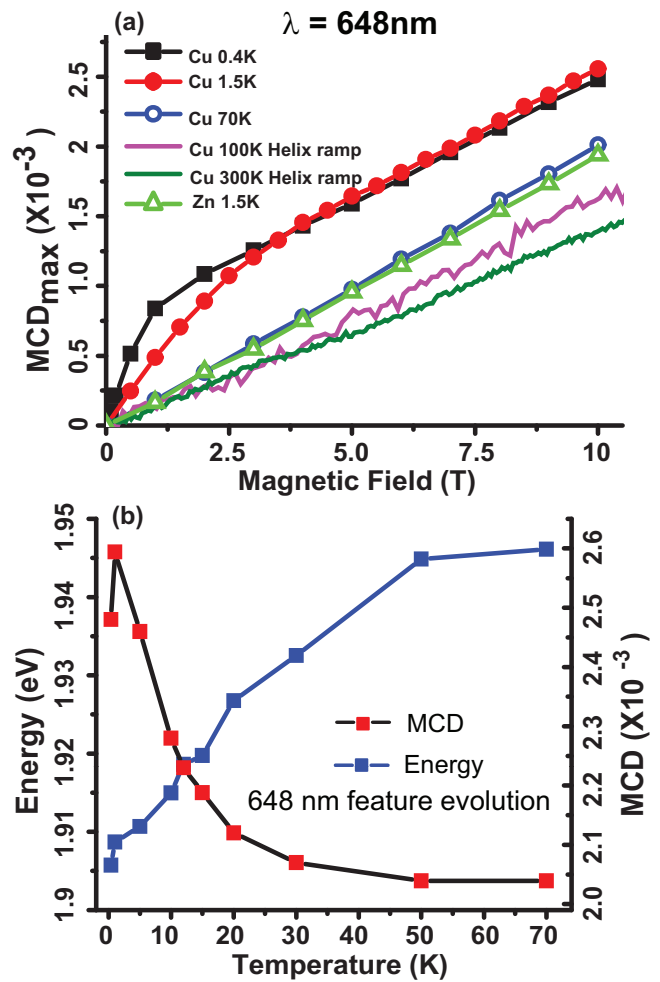}
	\caption{\label{fig:3} (Color online) (a) MCD ($ \lambda $ = 648 nm) evolution with increasing magnetic field in Cu-OBPc and Zn-OBPc at various temperatures ranging from 0.4 K to 300 K. For temperatures larger than 70 K MCD was measured in a continuous field sweep from 0 T to 25 T Split-Florida Helix magnet. (b) Evolution of the $ \lambda $ = 648 nm MCD with temperature for Cu-OBPc}
\end{figure}

\section{\label{sec:aed}\boldmath$\sigma\pi - d$ Exchange Interaction and enhanced effective \boldmath$g$ - factors}
 The results of the MCD experiments presented so far, clearly point towards the existence of an indirect exchange mechanism between the localized \textit{d}-shell ions, mediated by the delocalized $\pi$ electrons in the crystal, an organic analogue of the RKKY interaction where the $\sigma\pi-d$ exchange plays the role of the \textit{sp-d} interaction. It also identifies the specific electronic state within the bandgap manifold that is participating in this exchange. 
 A quantitative estimation for the strength of this $\sigma\pi-d$ exchange can be therefore obtained from the evolution of the Zeeman splitting with applied magnetic field, as measured in MCD experiments \cite{Furdyna1988,Bussian2009,Kuno1998}. In a straight forward interpretation of the original molecular field model for the \textit{sp-d} exchange, the hamiltonian will now take the form:
 \begin{equation}
 H_{ex} = \vec{S_{e}}\langle \vec{S}_{Cu}\rangle \sum\limits_{\vec{R}} J^{\sigma\pi-d}(\vec{r}-\vec{R})
 \end{equation}
where $\vec{S_{e}}$ is the delocalized electron spin, $\langle \vec{S}_{Cu}\rangle$ is the thermal average of the Cu ion spin, $J^{\sigma\pi-d}$ is the electron-ion exchange coupling constant and, $\vec{r}$ and $\vec{R}$ are coordinates of the delocalized electron and Cu$^{2+}$ ion respectively. In a perfect analogy to the DMS case, this will lead to a non-linear evolution of the delocalized electron Zeeman splitting with magnetic field which can be expressed as: 
\begin{equation}
\Delta E_{z} = g_{eff} \mu _B B
\end{equation}
where, $g_{eff}$ is an effective \textit{g}-factor proportional to the exchange integral and the total magnetic ion spin, $\mu _{B}$ is the Bohr magneton and B is the applied magnetic field. In II-Mn-VI DMSs, these effective \textit{g}-factors for both electrons and holes are readily extracted from MCD or magneto-photoluminescence experiments that measure exciton Zeeman splittings at low temperatures and high magnetic fields, and reach values as large as 100. Their sign is negative or positive depending on the ferromagnetic or antiferromagnetic nature of the \textit{sp-d} exchange.

Finding the effective \textit{g}-factor of the delocalized electrons in Cu-OBPc requires isolating the contributions of each MCD active state through a spectral fitting procedure, due to the broad spectral lineshapes.  The fitting procedure and assignment of states employed here relies on previous solid state MCD experiments of Hollebone and Dunford \cite{Huang1981,Keizer2003,Nyokong1987,Hollebone1978,Dunford1997} on thin film phthalocyanines, extensive literature about the nature and symmetries of molecular orbitals for the single CuPc molecule and an assumption that the degeneracies of certain molecular orbitals will be lifted in the crystal as a result of Jahn Teller effect and long range interactions present along the stacking axis. The three Q-band transition identified in the solution spectrum in Fig. 1 evolve into a seven-states manifold in the crystalline thin film. Contributions from these states to the MCD spectra are labeled 1 through 7 and color coded in Fig. 4. 

\begin{figure}[t!]
\centering
\includegraphics[width=\columnwidth]{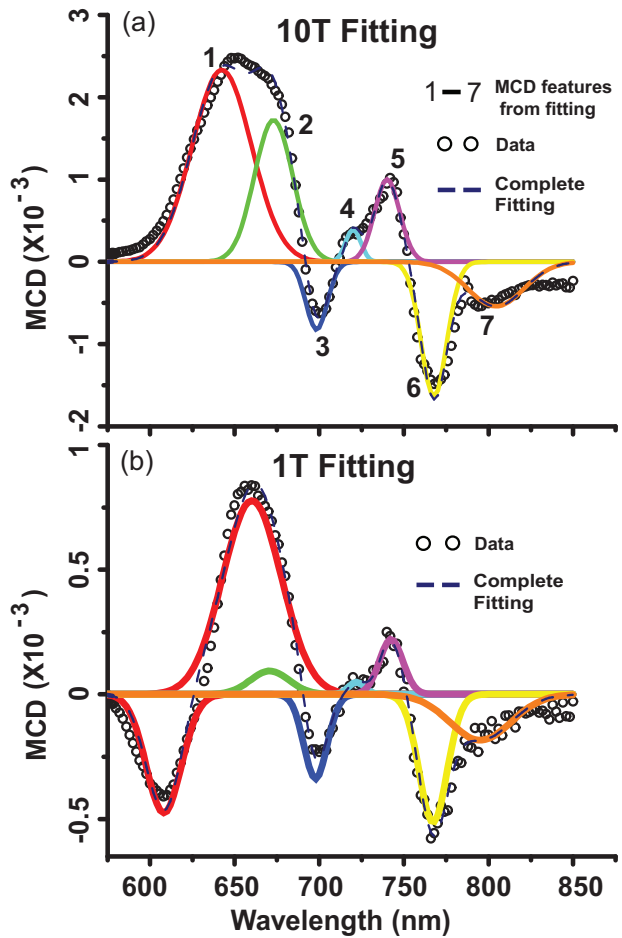}
\caption{\label{fig:4} (Color online) (a) 10 T Cu-OBPc MCD spectra gaussian fittings at 0.4 K. (b) 1 T Cu-OBPc MCD spectra gaussian fittings at 0.4 K }
\end{figure}

By comparing the high B-field spectra recorded at T = 300 K \cite{Pan2011}, and the T = 0.4 K (Fig. 4(a)), we can assign features 5 and 6 to the two HOMO-LUMO bandgap $a_{1u}(\pi) \rightarrow e_{2g}(\pi^{*})$ (Q$_{00}$) transitions. Similarly, features 2 and 3 represent the nominally forbidden $e_{g}(\pi) \rightarrow e_{1g}(\pi^{*})$ (Q$_{10}$) that are present in the MCD spectra because of the phonon coupling between the $a_{1u}$ and $e_{g}$ valence band states. Features 7 and 4 are phonon replicas of Q$_{00}$ and Q$_{10}$ respectively. Finally, feature 1 is associated with the optically allowed $e_{1u}(\sigma) \rightarrow e_{2g}(\pi^{*})$ (formerly Q$_{20}$) polarized along the stacking axis. In the low magnetic field regime (see B = 1 T fitting in Fig. 4(b)), the Q$_{20}$ region of the spectrum contains both negative and positive contributions, represented by the two red gaussians. The disappearance of the higher energy (negative MCD) contribution in high magnetic fields is the typical signature of a Zeeman-split degenerate ground state where the higher energy spin state is gradually depopulated as the magnetic field increases. All other features simply decrease in intensity without suffering any major line shape changes throughout the entire magnetic field range.


Figure 5(a) shows the magnetic field evolution of all the fitted transitions at 0.4 K as well as the 70 K fitted MCD data for feature 1. The distinct, Brillouin-like behavior of feature 1 that dominates the MCD spectrum at 0.4 K and the saturation of MCD for B $ > $ 2 T is now unambiguously resolved from the diamagnetic contributions of nearby states. The other features also display small deviations from the diamagnetic linear behavior in the small B-field regimes that most likely originate from the phonon- assisted mixing of $a_{1u}(\pi)$, $e_{1g}(\pi)$ and $e_{1u}(\sigma)$ states in the valence band manifold.   

 Quantitative information about the exchange can be extracted from MCD experiments through the effective \textit{g}-factor mentioned earlier. In this case, the Zeeman splitting of the $e_{1u}(\sigma)$ state is extracted from the fitted MCD lineshape of feature 1 and the corresponding fitted absorbance spectrum ( see figure S2 \cite{SuppInfo}), taking into account that the fullwidth at half maximum (FWHM) is larger than the Zeeman splitting, the latter is given by:
 \begin{equation}
 \Delta E_{z} = -2 \cdot \sigma \cdot \Delta A_{max}/A_{max}
 \end{equation}
 where, $\sigma$ is the feature 1 gaussian FWHM, $A_{max}$ is the maximum absorbance at $\Delta$$A_{max}$ which is the peak MCD signal of this feature \cite{Bussian2009,Kuno1998}. Because MCD measures the change in angular momentum associated with an electronic transition (i. e. the formation of an exciton), the measured effective \textit{g}-factor represents the sum of the \textit{g}-factors for the initial and final states involved in this transition (\textit{i. e.} the \textbf{exciton} \textit{g}-factor). In most organic molecules, electron \textit{g}-factors are typically equal to -2 for all orbitals. This is still the case for the $e_{g}(\pi^{*})$ conduction band states in CuPc since the unpaired Cu$^{2+}$ spin is expected to exclusively interact with states in the valence band manifold. Any observed enhancement in the measured \textit{g}-factors will therefore be the result of exchange interactions with the valence band states. 
 In Fig. 5(b), the evolution of this Zeeman splitting with magnetic field is plotted for the feature 1. An effective \textit{g}-factor of -6 is extracted from the low magnetic field slope, which results in a -4 valence band electron \textit{g}-factor. While this enhancement is not as spectacular as the one observed in inorganic DMSs, it is consistent with the relatively low ($ < $ 2 K) magnetic ordering temperatures measured in CuPc. More importantly, it holds great promise for future studies of other species such as CoPc and MnPc, where magnetization studies already indicate the presence of a much more robust magnetic order and ferromagnetism. 
\begin{figure}[b]
	\centering
	\includegraphics[width=8.6cm]{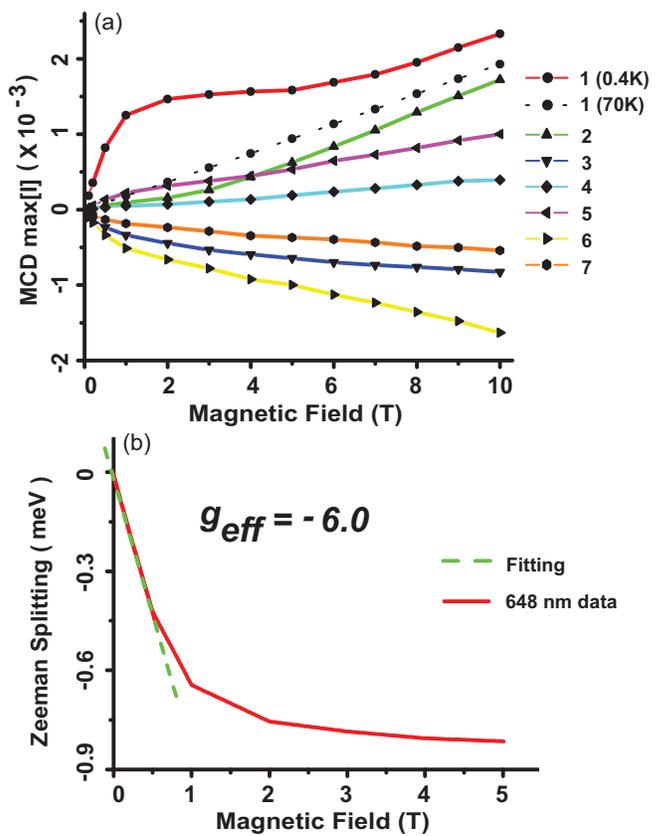}
	\caption{\label{fig:5} (Color online) (a) MCD evolution for the seven states with increasing magnetic field, only the feature 1 shows saturation at 0.4 K (b) Feature 1 Zeeman splitting evolution with magnetic field showing linear fitting for the effective \textit{g} factor}
\end{figure}

 A visual representation of the exchange mechanism between the delocalized ligand orbitals and energetically close unpaired \textit{d}-shell electron spin, as it is emerges from the low MCD findings reported in this paper is presented in Figure 6. In the absence of a full-fledged  bandgap structure calculation for Cu-OBPc, this representation employs the orbital symmetry notations for electronic states of single molecules with the assumption that long range intermolecular interactions along the stacking axis and crystal symmetry will mix the states located energetically close to the HOMO level. The $\sigma(\pi)-d$ exchange will predominately involve the former $e_{1u}(\sigma)$ orbitals of the ligand that are now mixed with the $\pi$ states and delocalized over distances larger than the nearest neighbor distance between the unpaired Cu spin located on the $b_{1g}$ \textit{d}- orbital.
Theoretical predictions by Wu et al. \cite{Wu2008,Wu2011}, that also rely on the molecular orbitals picture of the electronic states, point towards the $e_{1g}(\pi)$ orbital in the valence band as the state mediating the exchange mechanism. The evolution of MCD with magnetic field for each of these states, indicates, however, that the mechanism also involves the $e_{1u}(\sigma)$ orbitals which most likely hybridize with the $\pi$ states in the crystalline phase. Most importantly it indicates that one can take advantage of the large transition dipole of the optically -allowed $e_{u}(\sigma) \rightarrow e_{g}(\pi^{*})$ transition to optically create spin polarized electrons. 
 \begin{figure*}[t!]
 	\centering
 	\includegraphics[width=13cm]{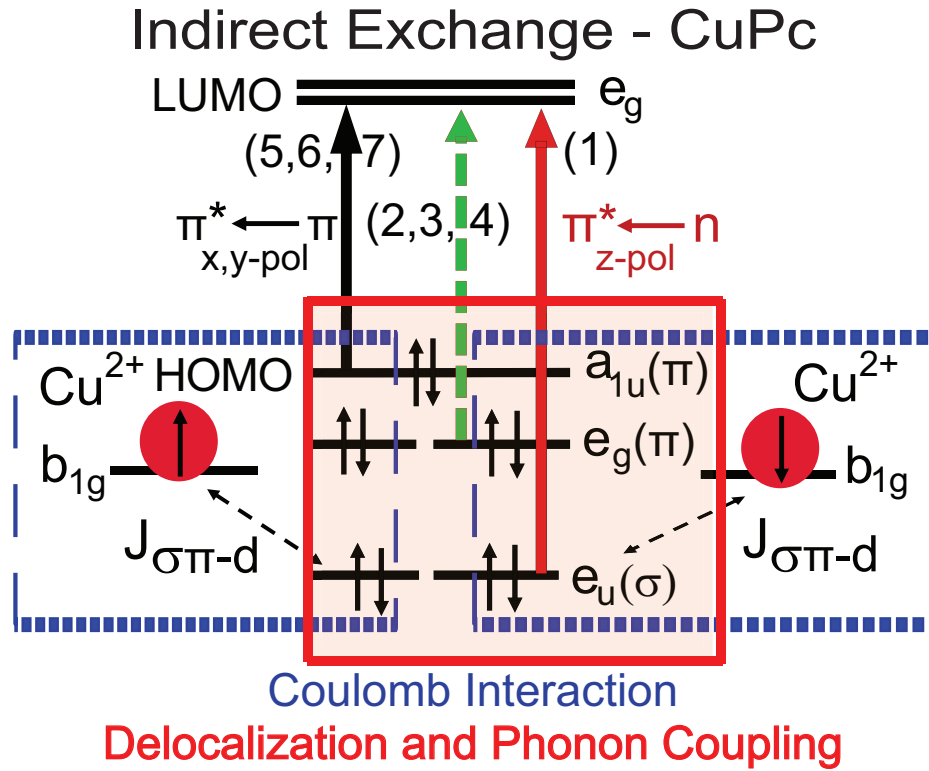}
 	\caption{\label{fig:6} (Color online) Schematic representation of the indirect exchange mechanism in Cu-OBPc. The Cu ion spins polarize delocalized electron in the $e_{g}/e_{u}$ states through Coulomb interaction. In turn these electrons align the nearest neighbor Cu spin anitferromagnetically. The solid arrows mark allowed transitions while dashed arrows indicate optically/symmetry forbidden transitions}
 \end{figure*}
\section{\label{sec:ccl} Conclusion}
\indent We performed variable temperature, variable field MCD Spectroscopy studies of Cu-OBPc  polycrystalline thin film for temperatures ranging 0.4 K to 300 K and magnetic fields up to 10 T. Evolution of seven MCD -active states identified in the bandgap region of the absorption spectrum was monitored as a function of temperature and magnetic field using a spectra fitting routine. One of these states, originating in the $\sigma$ orbitals of the lone nitrogen pair of the Pc ring bears the Brillouin-like signature of spin exchange with the magnetically ordered \textit{d}-shell unpaired Cu spin system. In analogy to the DMS molecular-field model of a temperature and magnetic field dependent electron \textit{g}-factor, the strength of the exchange mechanism between the delocalized ligand electrons and the \textit{d}-shell localized spins is reflected in an enhanced \textit{g}-factor of -4 at low temperatures for electrons occupying a specific ligand orbital. This study directly and unambiguously identifies the delocalized electron state responsible for mediating the indirect exchange mechanism that orders the Cu spins at low temperatures. 

From a broader perspective, these measurements constitute an interdisciplinary approach to probing magnetism in organic semiconductors that bridges quantum chemistry to condensed matter physics. Small molecule semiconductors such as the CuPc investigated here belong to an interesting intermediate regime where electrons are neither completely localized nor quasi-free and a hybrid model that accounts for longer range interactions and the role played by phonon coupling in the organic crystalline phase is perhaps necessary to fully describe all aspects of electronic and magnetic behavior. Many of these interesting phenomena were previously hidden by the overwhelming defects and disorder traditionally omnipresent in organic thin films. This problem is now circumvented by a whole new range of deposition techniques that improve the longe range ordering in films to a large degree. The studies presented here can be employed in the future as a feedback tool for designing and controlling this exchange interaction between conduction electrons and magnetic ion spins in the large family of porphyrins, which, in turn, could potentially lead to an organic spin injector material with realistic potential for a practical device, such as a spin valve. 
\section{\label{sec:exp} Experimental Section}
\indent Copper(II) 1,4,8,11,15,18,22,25-octabutoxy-29H,31H-phthalocyanine (Cu-OBPc) powder was purchased from Sigma-Aldrich and further purified by recrystallization in THF and ethanol \cite{Stevenson2003}. Metal-free OBPc powder was also purchased from Sigma-Aldrich and recrystallized in acetone. The purified OBPC crystals were used to synthesize the zinc(II)1,4,8,11,15,18,22,25-octabutoxy-29H,31H-phthalocyanine (Zn-OBPc) in-house, with slight modifications to the known procedure \cite{Mckeown1990}. The reaction was then followed by column chromatography and recrystallization in THF and Ethanol. 

Both of these Pc derivatives show increase in solubility in common organic solvents after purification. Crystalline thin films were deposited on \textbf{c}-plane cut sapphire (MCD-inactive) substrates using a rectangular capillary hollow pen writing technique developed at the University of Vermont by Headrick et al. \cite{Headrick2008,Wo2012}. The sapphire substrates were pretreated prior to thin film deposition by sonication in toluene, followed by methanol. A solution of purified Cu-OBPc dissolved in toluene or Zn-OBPc dissolved in THF with a concentration of 0.75 wt.$\%$ was loaded into the capillary where it was held in place capillary forces.  The sapphire substrate was mounted on a computer-controlled linear translation stage and film deposition was accomplished at room temperature by allowing the droplet of solution on the end of the capillary to make contact with the pretreated sapphire surface and then laterally translating the substrate at a controlled speed. The optimum writing speed used was 0.01 mm/s. For transmission geometry measurements, this deposition technique produces thin films with uniform thickness that ranges between 50 nm and 100 nm and grain sizes up to  a millimeter by appropriate control of the deposition parameters, ideally suited for spectroscopy techniques. The thickness of thin films and size of long grains can be optimized by varying the solution concentration and writing speed.

MCD and Absorption measurements were carried out with the sample mounted in a 10 T Oxford superconducting  magnet (Spectromag) equipped with a He3 insert for reaching sub-kelvin sample temperatures. The quasi -monochromatic output (bandwidth = 2 nm) of an Oriel 1/4 meter monochromator equipped with a 300 Watt xenon lamp was modulated into left and right circularly polarized light using a piezo-elastic modulator at a frequency of 50 kilo Hertz (V$_{ac}$) while an optical chopper operating at 266 Hertz measured the background signal (V$_{dc}$) \cite{Hipps1979}. Light was focused onto the sample using free space optics in Faraday geometry while the monochromator scans the wavelength through the UV-VIS-NIR range in steps of 1 nm and the transmitted light was collected by a silicon photodiode. MCD ($\Delta A \sim V_{ac}/V_{dc}$) and the overall transmittance (V$_{dc}$) were recorded using standard lockin amplifier techniques. The cryostat probe allowed for varying the sample temperature from 0.4 K to 80 K. For temperatures larger than 100 K we reproduced the experiment in the 25 T Split-Florida Helix magnet at the National High Magnetic Field Laboratory equipped with a nitrogen cooled custom sample holder. 


\section{\label{sec:ack} ACKNOWLEDGEMENTS} 
This work was supported by the National Science Foundation, Division of Materials Research MRI, CAREER and EPM program awards: DMR- 0722451, DMR- 0821268, DMR-1307017 and DMR-1056589, DMR-1229217. A portion of this work was performed at the National High Magnetic Field Laboratory which is supported by the National Science Foundation through DMR-1157490. We would like to especially thank Kelvin Chu from the Univ. of Vermont (UVM) Physics Dept and his former student Jacob Whalen Strothmann for many productive discussions on MCD spectral fittings for porphyrins and phthalocyanines. We are grateful to Lane Wright Manning (UVM-Materials Science) for his help with magnet cooldown during 5T preliminary characterization experiments.



\end{document}